%% file: main.tex
\documentclass{article}
\usepackage[utf8]{inputenc}
\usepackage{hyperref}
\usepackage{graphicx}
\usepackage{biblatex}
\usepackage{float}
\usepackage[margin=1.25in]{geometry}
\usepackage{setspace}
\title{Exploring Gender and Race Biases in the NFT Market}

\author{
  Howard Zhong\\
  MIT CSAIL \\
  \texttt{howardzh@mit.edu}
  \and
  Mark Hamilton\\
  MIT CSAIL, Microsoft \\
  \texttt{markth@mit.edu}
}

\date{}

\begin{document}
\maketitle

\begin{abstract}
Non-Fungible Tokens (NFTs) are non-interchangeable assets, usually digital art, which are stored on the blockchain. Preliminary studies find that female and darker-skinned NFTs are valued less than their male and lighter-skinned counterparts. However, these studies analyze only the CryptoPunks collection. We test the statistical significance of race and gender biases in the prices of CryptoPunks and present the first study of gender bias in the broader NFT market. We find evidence of racial bias but not gender bias. Our work also introduces a dataset of gender-labeled NFT collections to advance the broader study of social equity in this emerging market.
\end{abstract}


\input{introduction}

\input{method}

\input{results}
\input{discussion}
\input{conclusion}
\section{Acknowledgements}
This work is supported by the MIT Advanced Undergraduate SuperUROP program and supported by the National Science Foundation under Grant DGE-1747486. Any opinions, findings, and conclusions or recommendations expressed in this material are those of the author(s) and do not necessarily reflect the views of the National Science Foundation.

\printbibliography
\newpage

\input{appendix}

\end{document}

%% file: introduction.tex
\section{Introduction}

Are darker-skinned Non-Fungible Tokens (NFTs) sold for less than lighter-skinned NFTs? Or, are female NFTs worth less than male NFTs?  Egkolfopoulou and Gardner \cite{bloomberg} raises diversity concerns by reporting preliminary results on price differences based on gender and race. Determining whether diversity issues are present in the early-stage NFT space is important to raise awareness and bring change.

NFTs are tokens on a blockchain to represent ownership of a unique item. While the item can range from music, videos, or tweets, it usually represents the ownership of art. NFT's popularity has boomed since 2021 as the public became accustomed to the concept. We analyze NFTs stored on the Ethereum network \cite{ethereum2014whitepaper}

As NFTs are such a new concept, the dynamics of the NFT market has not been well studied. There has been previous work about NFT economics with focus on the trade networks \cite{nadini2021mapping}, the predictability of NFT sales \cite{nadini2021mapping}, and the risk and returns of NFT investments, \cite{mazur2021non}. Other topics of interest include how NFT art may change the dynamics of art markets \cite{bsteh2021painting}, how NFTs can affect game development by making the art tradable \cite{fowler2021tokenfication}, and how it can revolutionize digital ownership in different industries \cite{popescu2021non}. Among the studies, one specific collection that is most analyzed to represent the NFT market is CryptoPunks \cite{kong2021alternative} \cite{schaar2022non} \cite{das2021understanding}.  

\begin{figure}[!h]
    \centering
    \includegraphics[scale=.8]{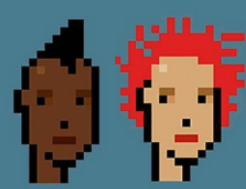}
    \caption{Dark Male CryptoPunk (left) and Light Female CryptoPunk (right). }
    \label{fig:punkteaser}
\end{figure}

The CryptoPunk collection is one of the most well-known NFT collections and consists of 10,000 avatars on the Ethereum blockchain created by Larva Labs in 2017. They are commonly credited for starting the NFT boom in 2021 \cite{upson_2021}. As shown in Figure \ref{fig:punkteaser}, CryptoPunks are avatars that can be used to represent oneself in the digital world and therefore are of interest when studying the NFT diversity problem.

We are interested in whether racial bias or gender bias exists in the NFT art market. In the December 2021 Bloomberg article ``Even in the Metaverse, Not All Identities Are Created Equal", Egkolfopoulou and Gardner \cite{bloomberg} examined graphs of sales prices of CryptoPunks over time across different genders and different races. They found that female avatars were sold for less than male avatars, and darker-skinned avatars were sold for less than lighter-skinned NFTs. However, their evidence for gender and racial biases is limited to visual inspection of CryptoPunk's historical price graphs instead. Thus, we test the statistical significance of race and gender biases in the prices of CryptoPunks and present the first study investigating gender bias in
the broader NFT market.

Our first goal is to analyze the gender and race biases of CryptoPunks rigorously through hypothesis testing. This will indicate whether price differences are significant enough to not be attributed to random chance. We find that darker-skinned CryptoPunks are sold for more than lighter-skinned CryptoPunks, but contrary to the analysis in \cite{bloomberg}, the difference in price for male and female CryptoPunks was not statistically significant. 

Our second goal is to extend our analysis to other collections to determine whether there are gender  biases for the broader NFT market. To that end, we analyze $44$ NFT collections labeled with gender tags and apply a hypothesis test to each collection to determine if male-looking NFTs sell at higher prices than female-looking ones. Like in CryptoPunks, we conclude that the price difference in male and female is not statistically significant. Because of the lack of race labels, we were only able to find $3$ collections to categorize between light and dark skinned NFTs, with analysis suggesting lighter-skinned NFTs are sold for more than darker-skinned NFTs.

%% file: method.tex
\section{Methodology} \label{methodology}
We describe the methods of analyzing the gender and race biases in the prices of NFTs. We first summarize our data collection process (Section \ref{methods-data} and \ref{methods-race-gender-data}) and then describe how we statistically quantify the gender and racial biases among different NFTs (Section \ref{methods-analysis}). The steps are shown in Figure \ref{fig:methodsfig}. More detailed description of methods and implementation can be found in the Appendix.

\begin{figure}[!h]

    \centering
    \includegraphics[scale=0.45]{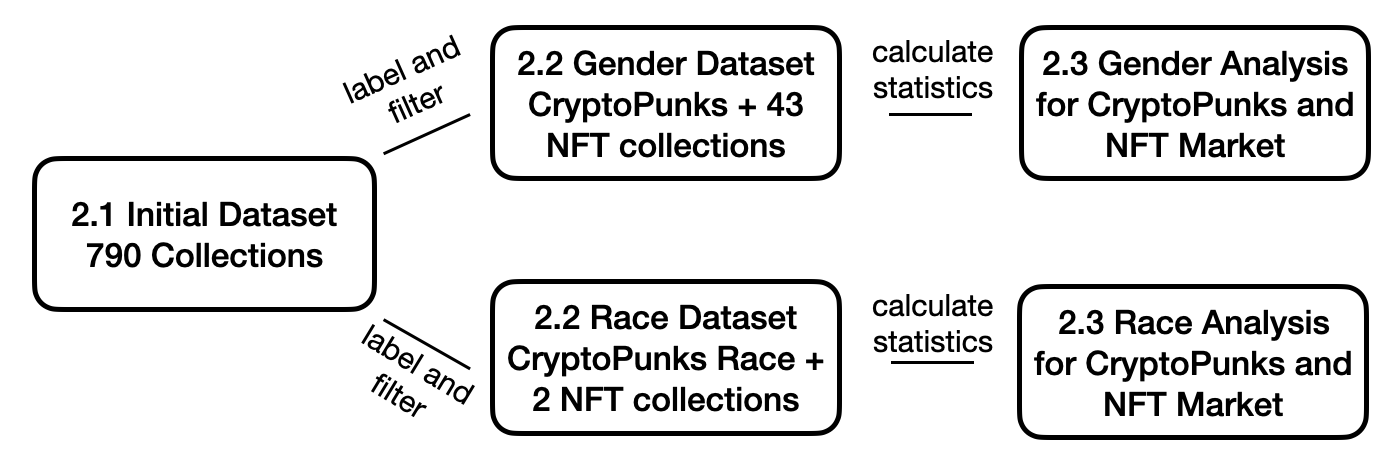}
    \caption{Flow Chart of Experiments.} 
    
    \label{fig:methodsfig}
\end{figure}

\subsection{Initial Data Collection} \label{methods-data}
Our dataset consists of NFTs transacted on OpenSea, which is the primary marketplace for NFTs on Ethereum. We query the OpenSea ``v1/collections" endpoint at the end of November 2022 \cite{openseaCollections} to retrieve NFT metadata and last sale price. We choose 790 collections from the Kaggle Ethereum NFTs dataset \cite{kaggleNFT} and NFTs from the top 30-day and all-time OpenSea volume leaderboard around November 2021. After the data collection process, we end up with $\sim 2.5$ NFTs, each that have been transacted upon.

\begin{table}[!h]
\centering
\begin{tabular}{|l|l|}
\hline
Number of Collections                                                              & 790         \\ \hline
Number of NFTs                                                            & 2.5 million \\ \hline
\begin{tabular}[c]{@{}l@{}}Number of Collections\\ with Gender Labels\end{tabular} & 44          \\ \hline
\end{tabular}
\caption{Summary of Dataset Collection}
\end{table}

\subsection{Retrieving Race and Gender Labels}
\label{methods-race-gender-data}
Many NFT collections do not represent humans, and therefore cannot be directly studied through the lens of race and gender. We select collections that have metadata with the words ``male" and ``female" and end with a total of $44$ such collections with gender labels representing different avatars. Statistics on these $44$ collections with gender labels can be found in Table \ref{table-collections} in Appendix A.2.

As far as we know, this is the first NFT dataset with gender labels across many collections. However, race labels often do not exist in the metadata, so we are limited to only CryptoPunks, Avastar, and Dynamic Duelers for collections with race labels. 

\subsection{Statistical Tools to Analyze Bias in Gender and Race}
\label{methods-analysis}
To determine the statistical significance of the hypothesis that female NFTs are sold for less than male NFTs, we run both paired and unpaired one-sided Student's t-tests \cite{walpole1993probability}. 

\textbf{Unpaired vs Paired t-test:} 
For unpaired t-test, we compare the mean of all NFT sale prices for male versus for female. For paired t-test, we calculate t-statistics on the paired difference of male and female prices marked to the daily mean price and the weekly mean price of the NFT. The paired t-test is used to isolate the male versus female price difference by fixing price variation across time.

\textbf{Log Transformation:} As t-test assume normality of the data, we apply a log transformation to address this.  With rare NFTs worth significantly more than common NFTs, NFT price distributions tend to follow a power law distribution \cite{nadini2021mapping}. Inspired by how stock prices follow a log-normal distribution \cite{antoniou2004log}, we applied the same transformation and found the distribution of log of price to be more normal. We refer to running the t-test on the log of prices as log t-test.

\textbf{Outlier Trimming:} Outliers may occur due to very high selling prices for rare NFTs or very low selling prices due to humans errors while listing. We address outliers using Winsorization \cite{tukey1963less}, or trimming outliers past a certain percentile. We report results for $0.1\%$, $1\%$, $2.5\%$, and $5\%$ percentiles. 

The approach described above is also used to compare the prices of light and dark CryptoPunks. We report results from combinations of different t-tests and outlier detection methods to show our conclusion remains consistent regardless of the way we conduct the statistical significance test. For the figures and statistics in this paper, unless otherwise stated, we remove outliers at the $2.5\%$ percentile.

%% file: results.tex
\section{Results}

Our goal is to investigate biases in pricing with respect to gender and race for NFT collections. We first analyze CryptoPunks, as it is one of the first NFT collections and is commonly credited for starting the 2021 NFT revolution \cite{upson_2021}. The dataset we use is compiled from querying the OpenSea API as described in Section \ref{methods-data}. 

\subsection{CryptoPunks}
While \cite{bloomberg} identifies biases in prices for both gender and race for CryptoPunks, we find only darker-skinned CryptoPunks are sold for less than lighter-skinned CryptoPunks.

\subsubsection{Gender} 

\begin{figure}[!h]
    \centering
    \includegraphics[scale=0.60]{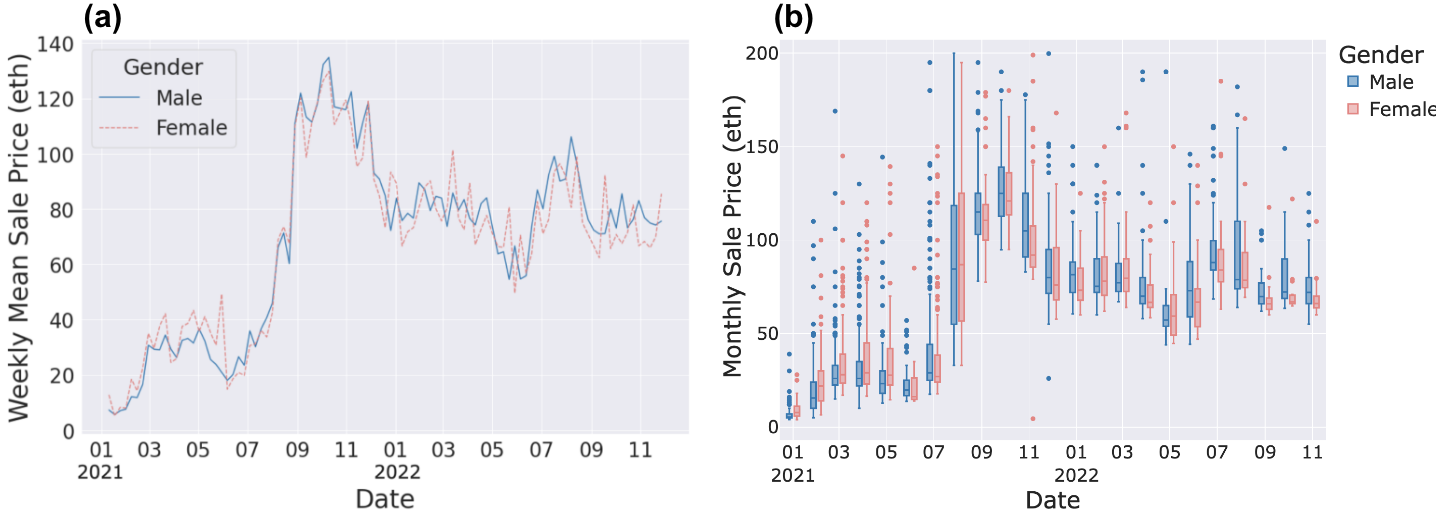}
    \caption{(a) Weekly Mean Sale Price and (b) Box plot of Monthly Sale Price for Male and Female CryptoPunks between 2021-01 and 2022-11.}
    \label{fig:cryptopunk-gender}
\end{figure}

We found price difference across gender was not significant. 
In Figure \ref{fig:cryptopunk-gender}a, male and female weekly mean sale price are roughly equal with blue and red lines being roughly level. This conclusion is corroborated by the male and female box plots at similar price levels for different months in Figure \ref{fig:cryptopunk-gender}b. 

We supplement the visual analysis with statistical anylsis. From 2021-01 to 2022-11, the median selling price of male CryptoPunks ($64.9$ eth) is greater than median price of female CryptoPunks ($64.0$ eth), but the mean selling price of male CryptoPunks ($63.5$ eth) is less than that of female CryptoPunks ($63.9$ eth). Because the means and medians yield different conclusions, we also analyze the p-value of different one-sided t-tests on whether male prices are greater than female prices. As shown in Table \ref{table-genderpunks}, we vary outlier detection percentiles, whether to run unpaired or paired t-test, and whether to apply the log transformation. Only at a $0.1\%$ outlier percentile level do some of the t-tests imply male price is greater than female price. This is reasonable because from Figure \ref{fig:cryptopunk-gender}b, male outliers tend to be higher priced than female outliers. As a precondition for the t-test is the absence of outliers, 
t-tests at other outlier levels ($2.5\%$ and $5\%)$ may be more informative. Thus, there is not enough statistical evidence to conclude there is gender bias in CryptoPunks.

\begin{table}[!h]
\centering
\begin{tabular}{|l|r|r|r|}
\hline
Outlier Percentile                 & 0.1\%          & 2.5\%          & 5.0\% \\ \hline
Paired t-test (marked to day)      & \textbf{0.027} & {0.719} & 0.803 \\ \hline
Paired t-test (marked to week)     & 0.103          & 0.815          & 0.778 \\ \hline
Unpaired t-test                    & 0.647          & 0.978          & 0.886 \\ \hline
Log Paired t-test (marked to day)  & \textbf{0.000} & {0.144} & 0.250 \\ \hline
Log Paired t-test (marked to week) & 0.114          & 0.240          & 0.129 \\ \hline
Log Unpaired t-test                & \textbf{0.005} & {0.623} & 0.549 \\ \hline
\end{tabular}
\caption{P-values varying outlier detection methods and types of one-sided t-tests for hypothesis that price of Male CryptoPunks $>$ Female CryptoPunks. \textbf{Bold} indicates p-value$<0.05$. }
\label{table-genderpunks}
\end{table}

\subsubsection{Race} 
Although we do not find gender bias in the pricing of CryptoPunks, we corroborate the visual analysis in \cite{bloomberg} with statistical results that lighter-skinned CryptoPunks are valued more than darker-skinned CryptoPunks.

\begin{figure}[!h]
    \centering
    \includegraphics[scale=0.58]{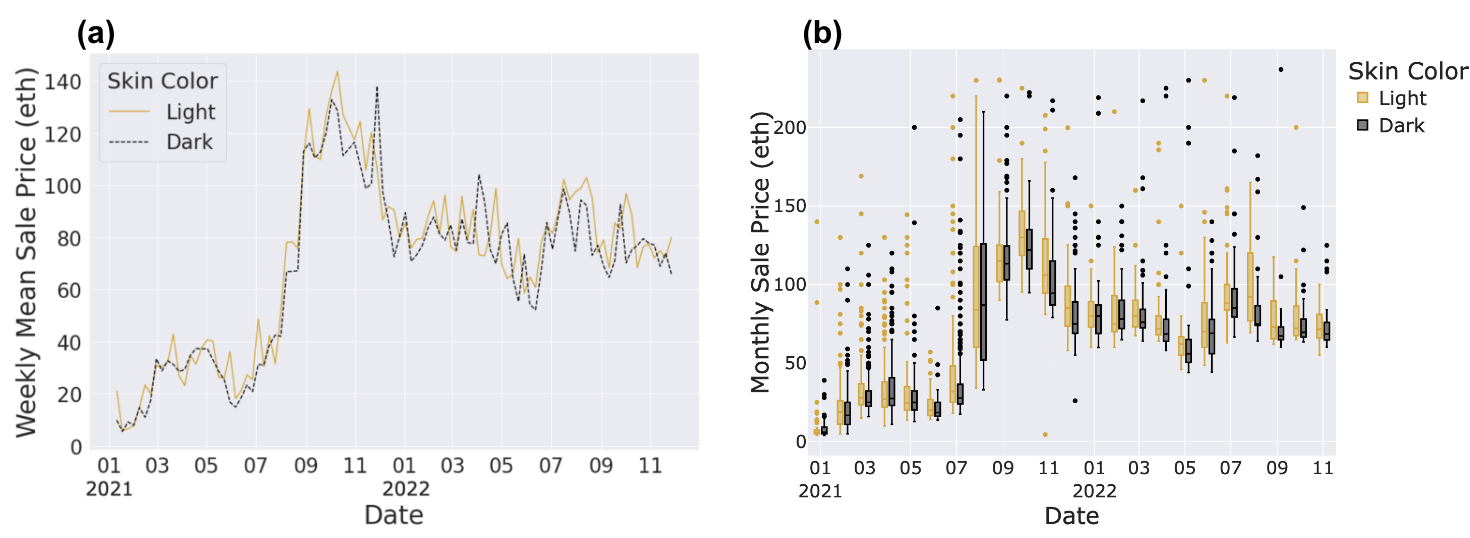}
    \caption{(a) Weekly Mean Sale Price and (b) Box plot of Monthly Sale Price for Light and Dark CryptoPunks between 2021-01 and 2022-11.}
    \label{fig:cryptopunk-race}
\end{figure}

We first plot the weekly mean sale prices of Dark and Light CryptoPunks between 2021-01 and 2022-11 in Figure \ref{fig:cryptopunk-race}a and find that Light CryptoPunks are consistently sold at a higher price than Dark CryptoPunks. Examining the box plot in Figure \ref{fig:cryptopunk-race}b for CryptoPunks prices across different months, we observe similar results. At a $2.5\%$ trim level, the median sale price for Light CryptoPunks ($65.6$ eth) is greater than that of Dark CryptoPunks ($64.0$ eth), and the mean sale price for Light CryptoPunks ($66.1$ eth) is also greater than that of Dark CryptoPunks ($64.5$ eth).
In addition, from Table \ref{table-racepunks}, every paired t-test across different outlier detection schemes supports this hypothesis with very low p-values, indicating the effect is large. Furthermore, the unpaired t-test results were above $0.05$ at $2.5\%$ and $5.0\%$ outlier trim levels may be due to CryptoPunks' large price variation across time that increases standard deviation and lowers t-stat. Thus, the evidence suggests that Light CryptoPunks are sold for more than Dark CryptoPunks.

\begin{table}[!h]
\centering
\begin{tabular}{|l|r|r|r|}
\hline
Outlier Percentile                 & 0.1\%             & 2.5\%             & 5.0\%             \\ \hline
Paired t-test (marked to day)      & \textbf{2.53E-04} & \textbf{2.00E-06} & \textbf{1.00E-06} \\ \hline
Paired t-test (marked to week)     & \textbf{6.00E-06} & \textbf{3.00E-06} & \textbf{2.72E-07} \\ \hline
Unpaired t-test                    & \textbf{3.46E-07} & 1.20E-01          & 1.17E-01          \\ \hline
Log Paired t-test (marked to day)  & \textbf{2.50E-07} & \textbf{2.81E-04} & \textbf{3.57E-04} \\ \hline
Log Paired t-test (marked to week) & \textbf{1.28E-07} & \textbf{4.50E-05} & \textbf{3.70E-05} \\ \hline
Log Unpaired t-test                & \textbf{1.01E-02} & {3.63E-01} & 2.39E-01          \\ \hline
\end{tabular}
\caption{P-values varying outlier detection methods and types of one-sided t-tests for hypothesis that price of Dark CryptoPunks $>$ Light CryptoPunks.  \textbf{Bold} indicates p-value$<0.05$}
\label{table-racepunks}
\end{table}

\subsection{Aggregate NFT Market}
We conduct similar analysis for gender and race bias across the NFT market. We found $44$ collections with gender labels but only $3$ with race labels.

\subsubsection{Gender} 
For the $44$ NFT collections chosen as detailed in Section \ref{methods-race-gender-data}, we calculate p-values to determine whether male NFT prices are greater than female NFT prices. P-values are calculated from a paired log t-test marked to weekly mean with a outlier trim percentile of $2.5\%$. To satisfy preconditions of a t-test, we apply a log transformation to make prices follow a more normal distribution and remove outliers. We mark to weekly mean price, which is common for low transaction volume markets such as NFT markets \cite{dowling2022fertile}. From the analysis, we find no clear relationship between market cap and gender bias. However, we do demonstrate that male prices are not statistically significantly higher than female prices for the broader NFT market.

\begin{figure}[H]
    \centering
    \includegraphics[scale=0.57]{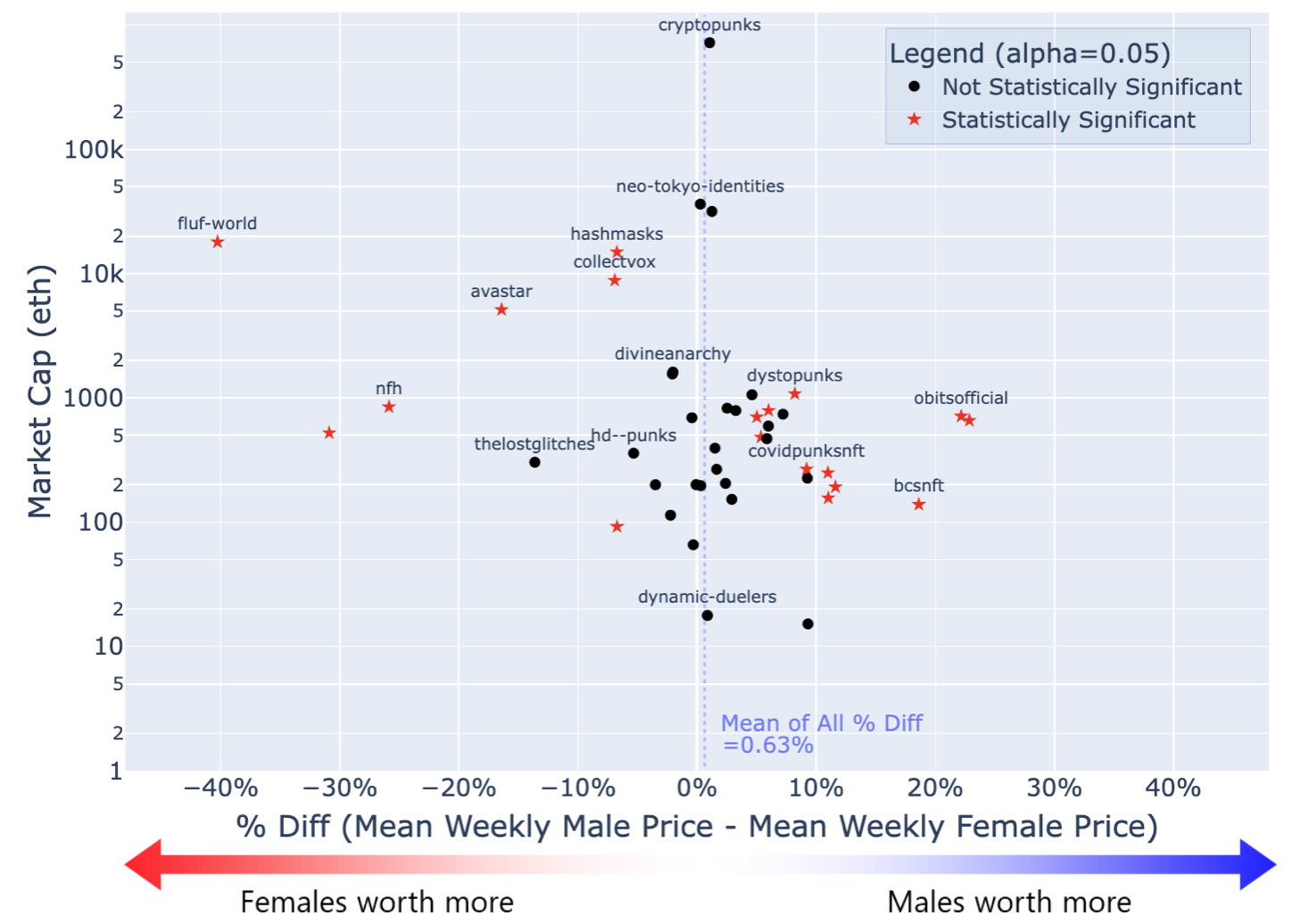}
    \caption{Plot of Market Cap versus $\%$ Mean of Weekly Difference in Male and Female Price per Collection. 
    }
    \label{fig:gender-price-marketcap-fig}
\end{figure}

Figure \ref{fig:gender-price-marketcap-fig} displays collections with male price statistically significantly ($\alpha=0.05$) higher than female price (right red dots), female price statistically significantly higher than male price (left red dots), and no statistically significant difference (black dots). 
There is no clear relationship between market cap and mean price difference across gender, but the figure does answer whether male NFTs are priced higher than female NFTs across the aggregate market. 

First, the mean across all collections of the price difference of male and female NFTs is only $0.63\%$. Furthermore, at a significance level of $\alpha=0.05$, only $11$ out of $44$ collections have male NFTs valued more than female NFTs, whereas $7$ collections have female NFTs valued more than male NFTs. Therefore, there is not enough evidence to conclude there is a statistically significant difference between male and female prices for the broader NFT market. 

\subsubsection{Race} 
In addition to CryptoPunks, we obtained race labels for the Avastar and Dynamic Duelers collections. From mean and median statistics in Table \ref{table-racestats}, we find that light-skinned NFTs on average tend to be sold for higher prices that darker-skinned counterpart. The result is less significant for Dynamic Duelers. For statistics in Table \ref{table-racestats} and \ref{table-racefull}, we use outlier trim level of $2.5\%$.

\begin{table}[H]
\centering
\begin{tabular}{|c|c|c|l|l|}
\hline
                 & Light Mean & Dark Mean & Light Median & Dark Median \\ \hline
avastar          & 0.235              & 0.229             & 0.219                & 0.2                 \\ \hline
dynamic\_duelers & 0.064              & 0.063             & 0.06                 & 0.06                \\ \hline
\end{tabular}
\caption{Mean and Median selling price (eth) of lighter-skinned and darker-skinned NFTs}
\label{table-racestats}
\end{table}

\begin{table}[H]
\centering
\begin{tabular}{|c|c|c|}
\hline
                                   & avastar & dynamic\_duelers \\ \hline
Log Paired t-test (marked to day)  & 0.158   & 0.190            \\ \hline
Log Paired t-test (marked to week) & 0.733   & 0.162            \\ \hline
Log Unpaired t-test                & \textbf{0.024}   & 0.573            \\ \hline
\end{tabular}
\caption{P-values varying types of one-sided log t-tests for the hypothesis that price of Lighter-skinned NFT $>$ Darker-skinned NFT. \textbf{Bold} indicates p-value$<0.05$.}
\label{table-racefull}
\end{table}

We also conduct hypothesis tests in Table \ref{table-racefull} to determine the statistical significance of whether lighter-skinned NFTs are worth more than darker-skinned NFTs. With a majority of the p-values less than $0.2$ and our conclusion from CryptoPunks, the results suggest lighter-skinned NFTs are worth more than darker-skinned NFTs, though not necessarily at a statistically significant level. Because this analysis is on a small subset of the entire NFT market, more data is required to generalize our conclusion to the entire NFT market. 

%% file: discussion.tex
\section{Discussion}
We seek to determine what mechanisms produce the different biases and why there exists racial bias but not gender bias in the prices of NFTs. Most NFTs in the market, especially those with gender and race attributes, represent avatars. These NFTs, especially expensive ones like CryptoPunks, can be used as profile picture, and thus people may prefer NFTs that look similar to themselves. We thus believe the demographics of NFT investors are the mechanisms that produce these biases. 

According to a data by Statistica \cite{best_2022} \cite{team_colormatics}, among the age group (18-34) with the largest interest in NFTs, men and women own NFTs fairly equally with 24\% men and 21\% women of total investor population. 
Thus, the fact we do not observe gender bias in NFT prices could be due to investors being roughly balanced across gender. 

However, the distribution of NFT buyers may be skewed white, which explains the existence of racial bias. Using Google searches as a proxy on the amount of NFT investors in a region, in the past five years, we found that the term ``NFT" was most searched in East Asia and North America and least searched in Africa and India. Thus, more lighter-skinned people are interested in NFTs compared to darker-skinned people and may be why lighter-skinned NFTs are more in demand. The full map of Google searches can be found in Figure \ref{fig:google-searches} in Appendix A.3. 

%% file: conclusion.tex
\section{Conclusion}
In this paper, we provide a rigorous statistical analysis of the gender and race biases of CryptoPunks and the NFT market as a whole. We found that gender bias is not statistically significant for CryptoPunks, but racial bias is. Regardless of how we remove outliers, whether we apply a log transform, or whether we used a paired or unpaired t-test, our conclusion remains consistent. As CryptoPunks are well-known and commonly credited for starting the rise of NFTs, biases in prices reflect early-investors' perceptions of NFTs representing different races. 

When we analyzed the NFT market as a whole, we also found there was not a statistically significant difference in price between male and female NFTs. For future work, we plan to label race for more NFT datasets to be able to conclude whether the trend of lighter-skinned CryptoPunks being sold for more than darker-skinned CryptoPunks holds for the general NFT market. We believe this price disparity may be due to the demographics of NFT investors. For avatars that are NFTs, people may tend to buy ones that look similar to their appearance.  

Investigating the gender and race biases are important from a culture standpoint due to the popularity of NFTs. As the metaverse is at its seed stage of creation with egalitarian values at the forefront of decentralization, it is important that inequity and biases in society do not propagate into it. Identifying these racial biases is the first step to drive initiatives that bring equity to NFTs and the metaverse. Possible countermeasures of improving fairness include raising awareness to the issue and increasing access to NFTs to other parts of the world, especially in developing countries. We hope future investors will purchase NFTs with racial sensitivity in mind. 

%% file: appendix.tex
\section*{A Appendix}

\subsection*{A.1 Implementation Details} \label{append-methods}
We elaborate on the methods and the implementation details. This sections follows the structure Section \ref{methodology}.



    

\subsubsection*{A.1.1 Initial Data Collection} \label{append-methods-data}
Our dataset only includes NFTs transacted on OpenSea, which is the primary marketplace for NFTs on the Ethereum blockchain. We query the OpenSea ``v1/collections" endpoint at the end of November 2022 \cite{openseaCollections} to retrieve collection metadata as well as each individual NFT's metadata and last sale price. We choose 790 collection names from the Kaggle Ethereum NFTs dataset \cite{kaggleNFT} and NFTs from the top 30-day and all-time OpenSea volume leaderboard around November 2021. Most collections have around 5,000 to 10,000 items, and our queries yield around 70-80\% of the data as the other 20-30\% do not have transactions.

After querying the Application Programming Interface (API) using a script written with the PySpark framework \cite{pyspark} on Databricks, we create a dataframe containing each NFT's collection name, id, sale information, metadata, and image url. We then loop through this dataframe with multiple workers to download the images in parallel. In the end, we obtain a dataset of about $\sim 2.5$ NFTs, each that have been transacted upon. 


\subsubsection*{A.1.2 Retrieving Race and Gender Labels}
\label{append-methods-race-gender-data}
To get gender-labelled data, we select collections that have metadata with the words ``male" and ``female" to find collections where at least $50\%$ of the images have gender labels. We filter out collections that have more than $75\%$ male or female to avoid rarity factors affecting the price. We find a total of $44$ such collections with gender labels that satisfy the criteria.

These collections with gender labels are

\begin{figure}[!h]
    \centering
    \includegraphics[scale=0.58]{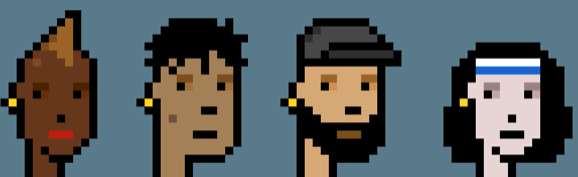}
    \caption{The 4 Different Shades of CryptoPunks. From left to right, the shades are darkest, mid, lighter, and lightest. We group darkest and mid as Dark and lighter and lightest as Light.}
    \label{fig:cryptopunk-race-ex}
\end{figure}

It is more challenging to determine the race of an NFT collection because most collections do not directly annotate this information. For the CryptoPunks collection, we label each NFT based on the frequencies of the four different pixel values representing the four different skin tones as displayed in Figure \ref{fig:cryptopunk-race-ex}. The above process was inspired by \cite{cryptopunksnotdead_2021}. For our analysis, we group darkest and mid as Dark and lighter and lightest as Light. For other collections, we were unable to retrieve race labels from the metadata.  

We also attempted to find race labels for a subset of the $44$ collections with gender labels with the criteria that there were a high proportion of items that could be categorized as either dark or light skinned. We filtered out collections that have more than $75\%$ light-skinned or dark-skinned to avoid rarity factors affecting the price. This required manual review because most collections do not have easy-to-search race labels in the metadata, or even way to categorize race.
For Avastar, we categorized ``Amber Brown" as dark-skinned and ``Pale Pink" as light-skinned in the ``skin\_tone" attribute. For Dynamic Duelers, we categorized ``Black" as dark-skinned and ``White" as light-skinned in the ``Skin color" attribute.
 
\subsubsection*{A.1.3 Statistical Tools to Analyze Bias in Gender and Race}
\label{append-methods-analysis}
We detail the statistical methods used to analyze the bias in NFT pricing for gender and race. Specifically, we go into more detail about the implementation and underlying assumptions of these tests.

\textbf{Unpaired t-test} assumes 
\begin{enumerate}
    \item independence of observations and each point belongs to either male or female
    \item the variances of male and female prices are equal for the population
    \item the prices follows a normal distribution 
    \item no significant outliers in each group

\end{enumerate}
The unpaired t-test is robust to all but large deviation from these assumptions. Assumption 1 and 2 are satisfied, but 3 and 4 require more careful consideration. 

\textbf{Log Transformation:} For assumption 3, with rare NFTs worth significantly more than common NFTs, NFT price distributions tend to follow a power law distribution \cite{nadini2021mapping}. Because it's well known that stock prices follow a log-normal distribution \cite{antoniou2004log}, we apply the same transformation and found log of price tended to be closer to a normal distribution. We thus decided to also run the t-test on the log of prices, which we refer to as log t-test.

\textbf{Outlier Trimming:} For assumption 4, as outliers may occur due to very high selling prices for rare NFTs or very low selling prices due to humans errors while listing, we address this issue by Winsorization \cite{tukey1963less}, or trimming outliers past a certain percentile. Specifically, we report t-test results while trimming both left and right tails of the distribution at $0.1\%$, $1\%$, $2.5\%$, and $5\%$ percentiles. 

\textbf{Paired t-test:} In addition to the unpaired t-test for independent samples, we also run a paired t-test for two dependent samples in order to better isolate the male versus female price difference while fixing time. Specifically, we calculate t-statistics on the paired difference of male and female prices marked to the daily mean price and the weekly mean price of the NFT. Assumptions 1 and 2 are now (1) male - female price must be continuous and (2) observations are independent of one another. Assumptions 3 and 4 from unpaired test remain the same and. We address assumptions 3 and 4 with the log-transformation and Winsorization, as described earlier.

The approach described above is also used to compare the prices of lighter-skinned and darker-skinned CryptoPunks. We utilize the SciPy Python package \cite{2020SciPy-NMeth} to implement both types of t-tests.

\subsection*{A.2 Detailed NFT Information}
The 44 collections that we sampled and analyzed are all NFTs representing avatars. In Table \ref{table-collections}, month created, mean sale price, median sale price, 2.5\% trim mean sale price, and total supply are listed. 

\begin{table}[H]
\centering
\begin{tabular}{|l|l|l|l|l|l|}
\hline
collection                 & month created & mean   & median & mean (trim 2.5\%) & total supply \\ \hline
avariksagauniverse         & 2021-09        & 0.15   & 0.10   & 0.11              & 8888          \\ \hline
avastar                    & 2020-02        & 0.45   & 0.20   & 0.27              & 25458         \\ \hline
bcsnft                     & 2021-09        & 0.09   & 0.08   & 0.08              & 7777          \\ \hline
bored-mummy-baby-waking-up & 2021-08        & 0.13   & 0.10   & 0.11              & 3888          \\ \hline
collectvox                 & 2021-07        & 1.85   & 0.60   & 1.14              & 8888          \\ \hline
covidpunksnft              & 2021-07        & 0.20   & 0.08   & 0.11              & 10000         \\ \hline
crypto-hodlers-nft         & 2021-07        & 0.14   & 0.07   & 0.08              & 10000         \\ \hline
cryptomutts-official       & 2021-09        & 0.12   & 0.06   & 0.08              & 10000         \\ \hline
cryptopunks                & 2017-06        & 78.41  & 33.00  & 45.88             & 10000         \\ \hline
currencypunks              & 2021-09        & 0.04   & 0.03   & 0.03              & 10000         \\ \hline
cyphercity                 & 2021-07        & 0.07   & 0.03   & 0.04              & 8886          \\ \hline
divineanarchy              & 2021-11        & 0.32   & 0.29   & 0.29              & 10011         \\ \hline
doobits                    & 2021-07        & 0.02   & 0.02   & 0.02              & 10024         \\ \hline
dynamic-duelers            & 2021-08        & 0.06   & 0.06   & 0.06              & 983           \\ \hline
dystopunks                 & 2021-06        & 2.39   & 1.10   & 1.87              & 2077          \\ \hline
evaverse                   & 2021-07        & 0.28   & 0.22   & 0.25              & 10000         \\ \hline
expansionpunks             & 2021-08        & 0.30   & 0.17   & 0.23              & 10000         \\ \hline
fluf-world                 & 2021-08        & 3.05   & 1.33   & 2.06              & 10000         \\ \hline
fourierpunks               & 2021-06        & 0.13   & 0.10   & 0.11              & 420           \\ \hline
guardians-of-the-metaverse & 2021-09        & 0.16   & 0.14   & 0.15              & 10000         \\ \hline
hashmasks                  & 2021-01        & 43.34  & 1.31   & 1.47              & 16384         \\ \hline
hd--punks                  & 2021-06        & 0.10   & 0.04   & 0.05              & 10000         \\ \hline
heavencomputer             & 2021-08        & 0.27   & 0.17   & 0.25              & 7777          \\ \hline
influence-crew             & 2021-09        & 0.24   & 0.10   & 0.15              & 7562          \\ \hline
lucha-libre-knockout       & 2021-08        & 0.05   & 0.02   & 0.02              & 10000         \\ \hline
metahero-generative        & 2021-09        & 5.69   & 5.25   & 5.44              & 6458          \\ \hline
misfit-university-official & 2021-06        & 0.05   & 0.02   & 0.03              & 10000         \\ \hline
mutant-punks-nft           & 2021-10        & 0.05   & 0.04   & 0.04              & 10000         \\ \hline
neo-tokyo-identities       & 2021-10        & 14.75  & 11.00  & 13.67             & 2021          \\ \hline
nfh                        & 2021-09        & 0.17   & 0.12   & 0.14              & 8888          \\ \hline
obitsofficial              & 2021-09        & 0.23   & 0.16   & 0.18              & 7132          \\ \hline
octohedz                   & 2021-08        & 2.05   & 1.48   & 1.83              & 888           \\ \hline
octohedz-reloaded          & 2021-10        & 0.10   & 0.07   & 0.09              & 8001          \\ \hline
pixls-official             & 2021-03        & 0.25   & 0.15   & 0.20              & 5468          \\ \hline
role-for-metaverse         & 2021-09        & 0.15   & 0.13   & 0.14              & 9800          \\ \hline
rug-wtf                    & 2021-07        & 0.06   & 0.03   & 0.05              & 10000         \\ \hline
skvllpvnkz-hideout         & 2021-09        & 0.11   & 0.08   & 0.09              & 10000         \\ \hline
spunks-nft                 & 2021-06        & 0.21   & 0.09   & 0.12              & 10000         \\ \hline
srsc                       & 2021-06        & 240.37 & 0.07   & 0.07              & 8888          \\ \hline
stoned-ape-saturn-club     & 2021-09        & 0.08   & 0.05   & 0.07              & 6969          \\ \hline
theantimasks               & 2021-04        & 0.06   & 0.06   & 0.05              & 1551          \\ \hline
thelostglitches            & 2021-08        & 0.18   & 0.09   & 0.13              & 9999          \\ \hline
wearetheoutkast            & 2021-09        & 0.07   & 0.07   & 0.07              & 9996          \\ \hline
zunks                      & 2021-08        & 0.16   & 0.08   & 0.11              & 10000         \\ \hline
\end{tabular}
\caption{Statistics for the 44 NFT collections analyzed.}
\label{table-collections}
\end{table}

\subsection*{A.3 Google NFT Searches Map}
\label{appendix-searches}
Below is a map of Google search data for the term ``NFT" in the past 5 years (as of January 2023). Most searches occur in East Asia, North America, and Australia. Europe, South America, Middle East, and India roughly have the same amount of searches. Africa has the least number of searches. 

\begin{figure}[!h]
    \centering
    \includegraphics[scale=0.88]{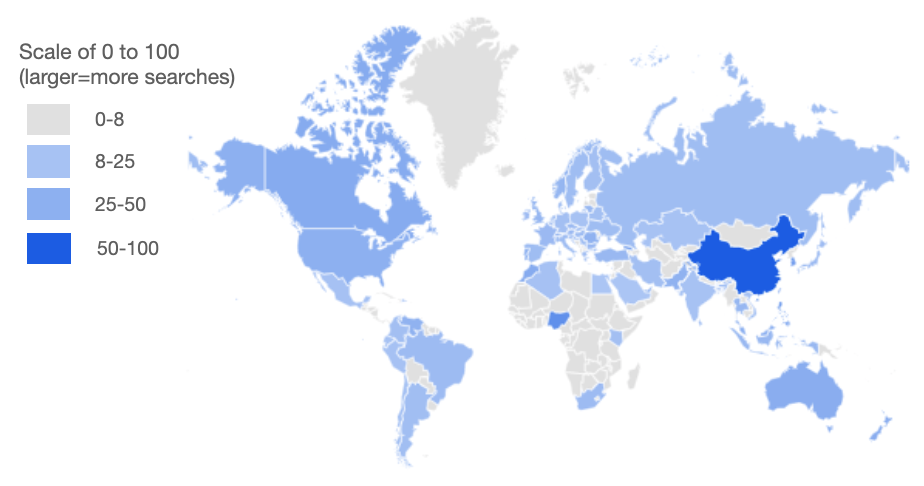}
    \caption{Google Search data for the term "NFT" for different countries}
    \label{fig:google-searches}
\end{figure}